\documentstyle[aps,pra,palatcm]{revtex}

\makeatletter
\renewcommand{\@seccntformat}[1]{\relax}
\renewcommand{\section}{\@startsection%
{section}{1}{0pt}{3.5ex \@plus 1ex \@minus .2ex}{2.3ex \@plus.2ex}%
{\raggedright\normalsize\bfseries\rmfamily}}  %
\renewcommand{\subsection}{\@startsection%
{subsection}{2}{0pt}{3.25ex \@plus 1ex \@minus .2ex}{1.5ex \@plus.2ex}%
{\raggedright\normalsize\itshape\rmfamily}}
\renewcommand{\subsubsection}{\@startsection%
{subsubsection}{3}{0pt}{3.25ex \@plus 1ex \@minus .2ex}{1.5ex \@plus.2ex}%
{\raggedright\normalsize\itshape\rmfamily}}
\renewcommand{\@seccntformat}[1]{{\csname the#1\endcsname}.\hspace{1em}}

\renewcommand{\p@section}{}

\renewcommand{\p@subsection}{}

\renewcommand{\p@subsubsection}{}

\makeatother

\newcommand{\I}{\mathrm{i}}
\newcommand{\lkl}{\left(}\newcommand{\rkl}{\right)}

\newcommand{\phpr}{^{\phantom{\prime}}}
\newcommand{\phbig}{\vphantom{\big|}}
\newcommand{\phBig}{\vphantom{\Big|}}

\newcommand{\half}{\frac{1}{2}}

\newcommand{\spinhalf}{\mbox{spin-$\half$}}
\newcommand{\spinone}{\mbox{spin-$1$}}
\newcommand{\bra}[1]{\left\langle{#1}\right|}
\newcommand{\ket}[1]{\left|{#1}\right\rangle}
\newcommand{\braket}[2]{\langle{#1}|{#2}\rangle}

\newcommand{\Exp}[1]{\mathrm{e}^{\mbox{\footnotesize$\textstyle{#1}$}}}
\makeatletter
\newcommand{\Text}[1]{%
{\mathchoice{\text@D{#1}}{\text@T{#1}}{\text@F{#1}}{\text@S{#1}}}}
\newcommand{\text@D}[1]{\mbox{\normalsize\textrm{#1}}}
\newcommand{\text@T}[1]{\mbox{\normalsize\textrm{#1}}}
\newcommand{\text@F}[1]{\mbox{\footnotesize\textrm{#1}}}
\newcommand{\text@S}[1]{\mbox{\scriptsize\textrm{#1}}}
\makeatother
\begin{document}

\title{The mean king's problem: Spin 1}

\author{Yakir Aharonov$^{a,b}$ and Berthold-Georg Englert$^{c,d}$}

\address{%
$^a$School of Physics and Astronomy, Tel-Aviv University, %
Tel-Aviv 69978, Israel\\
$^b$Physics Department, University of South Carolina, %
Columbia SC 29208, USA\\
$^c$Atominstitut der \"Osterreichischen Universit\"aten, Stadionallee 2, %
1020 Wien, Austria\\
$^d$Max-Planck-Institut f\"ur Quantenoptik, Hans-Kopfermann-Stra\ss{}e 1, %
85748~Garching, Germany}

\date{To appear in Zeitschrift f\"ur Naturforschung, received 26 November 2000}

\wideabs{
\maketitle\begin{abstract}%
We show how one can ascertain the values of four mutually complementary
observables of a \spinone\ degree of freedom.\\{}
\end{abstract}
}

\section{Introduction}\label{sec:Intro}
About a dozen years ago, one of us (YA) co-authored a paper \cite{VAA} 
with the somewhat provocative title
``How to ascertain the values of $\sigma_x$, $\sigma_y$, and $\sigma_z$
of a spin-$\frac{1}{2}$ particle''.
It reports the solution of what has later become known as 
\emph{The King's Problem}:
\begin{quotation}
A ship-wrecked physicist gets stranded on a far-away island that is
ruled by a mean king who loves cats and hates physicists since the day
when he first heard what happened to Schr\"odinger's cat.
A similar fate is awaiting the stranded physicist.
Yet, mean as he is, the king enjoys defeating physicists on their own
turf, and therefore he maliciously offers an apparently virtual chance of 
rescue.

He takes the physicist to the royal laboratory, a splendid place where
experiments of any kind can be performed perfectly.
There the king invites the physicist to prepare a certain silver atom in 
any state she likes.
The king's men will then measure one of the three cartesian spin components 
of this atom --- they'll either measure $\sigma_x$, $\sigma_y$, or $\sigma_z$
without, however, telling the physicist which one of these measurements is
actually done.
Then it is again the physicist's turn, and she can perform any experiment of
her choosing.
Only after she's finished with it, the king will tell her which spin component
had been measured by his men.
To save her neck, the physicist must then state correctly the measurement
result that the king's men had obtained.

Much to the king's frustration, the physicist rises to the challenge --- and
not just by sheer luck:
She gets the right answer any time the whole procedure is repeated.
How does she do it?
\end{quotation}

Readers who don't know the answer should try to figure it out themselves
rather than consult the said reference.
There is a lesson here about the wonderful things entanglement can do for you.

It is worth mentioning that this thought experiment of 1987 has not been
realized as yet.
Very recently, however, a quantum-optical analog has been formulated
\cite{EKW}, and it is hoped that experimental data will be at hand shortly.

The present paper deals with a generalization of the king's problem.
Instead of the traditional spin-$\frac{1}{2}$ atom, we consider the situation
of a spin-$1$ atom.
The two main questions are then:
What are the appropriate \mbox{spin-$1$} analogs of the spin-$\frac{1}{2}$
observables $\sigma_x$, $\sigma_y$, $\sigma_z$?
And, how does the physicist save her neck now?

The first question is answered in Sect.~\ref{sec:CSMCO} in terms of a
complete set of mutually complementary observables.
The answer to the second question is given in Sect.~\ref{sec:SaveNeck};
it employs essentially the same strategy that works in the \spinhalf\
case, so that we have a genuine generalization indeed.
Further generalizations to even higher spins will be discussed elsewhere
\cite{YA+BGE:ip}.

\section{Mutually complementary observables}\label{sec:CSMCO}
The three \spinhalf\ observables $\sigma_x$, $\sigma_y$, $\sigma_z$ are
\emph{complete} in the sense that the probabilities for finding their
eigenvalues as the results of measurements specify uniquely the statistical
operator that characterizes the \spinhalf\ degree of freedom of the ensemble
under consideration.
They are not overcomplete because this unique specification is not ensured if
one of the spin components is left out.

In addition to being complete, the observables $\sigma_x$, $\sigma_y$,
$\sigma_z$ are also pairwise \emph{complementary}, which is to say that in a
state where one of them has a definite value, all measurement results for the
other ones are equally probable.
For example, if $\sigma_x=1$ specifies the ensemble, say, then the results of
$\sigma_y$ measurements are utterly unpredictable: $+1$ and $-1$ are found
with equal frequency; and the same applies to $\sigma_z$ measurements.

What is essential here are not the eigenvalues of  $\sigma_x$, $\sigma_y$,
$\sigma_z$, but their sets of eigenstates.
It is familiar that they are related to each other by
\begin{eqnarray}
\ket{\sigma_x=\pm1}&=&
2^{-\half}\lkl\phbig\ket{\sigma_z=+1}\pm\ket{\sigma_z=-1}\rkl\,,
\nonumber\\
\ket{\sigma_y=\pm1}&=&
2^{-\half}\lkl\phbig\ket{\sigma_z=+1}\pm\I\ket{\sigma_z=-1}\rkl\,,
\label{eq:B1}
\end{eqnarray}
if the usual phase conventions are adopted.
The fact that the transition probabilities
\begin{eqnarray}
\biglb|\braket{\sigma_x=\pm1}{\sigma_y=\pm1}\bigrb|^2&=&\half\,,\nonumber\\
\biglb|\braket{\sigma_y=\pm1}{\sigma_z=\pm1}\bigrb|^2&=&\half\,,\nonumber\\
\biglb|\braket{\sigma_z=\pm1}{\sigma_x=\pm1}\bigrb|^2&=&\half\,,  
\label{eq:B2}
\end{eqnarray}
do not depend on the quantum numbers $\pm1$, is the statement of the pairwise 
complementary nature of $\sigma_x$, $\sigma_y$, and $\sigma_z$.
Their algebraic completeness is then an immediate consequence of the
insight that a \spinhalf\ degree of freedom can have at most three
mutually complementary observables \cite{WooFie}.

Analogously, there can be no more than four such observables for a \spinone\
degree of freedom.
Let's call them $A_0$, $A_1$, $A_2$, and $A_3$, and to be specific, 
we take their eigenvalues to be $0$, $1$, and $2$.
We denote by $\ket{m_k}$ the eigenstate of $A_m$ to eigenvalue $k$,
and we express the eigenstates of $A_1$, $A_2$, $A_3$ in terms of those of
$A_0$.
With
\begin{equation}
  \label{eq:B3}
  x=\Exp{\I2\pi/3}\,,
\end{equation}
the basic cubic root of unity, it is a matter of inspection to verify that the
choice 
\begin{eqnarray}
\bigl(\ket{1_0},\ket{1_1},\ket{1_2}\bigr)&=&
\bigl(\ket{0_0},\ket{0_1},\ket{0_2}\bigr)
\frac{1}{\sqrt{3}}\lkl\begin{array}{ccc}
x & 1 & 1 \\ 1 & x & 1 \\ 1 & 1 & x
\end{array}\rkl\,,
\nonumber\\
\bigl(\ket{2_0},\ket{2_1},\ket{2_2}\bigr)&=&
\bigl(\ket{0_0},\ket{0_1},\ket{0_2}\bigr)
\frac{1}{\sqrt{3}}\lkl\begin{array}{ccc}
x^2 & 1 & 1 \\ 1 & x^2 & 1 \\ 1 & 1 & x^2
\end{array}\rkl\,,
\nonumber\\
\bigl(\ket{3_0},\ket{3_1},\ket{3_2}\bigr)&=&
\bigl(\ket{0_0},\ket{0_1},\ket{0_2}\bigr)
\frac{1}{\sqrt{3}}\lkl\begin{array}{ccc}
1 & 1 & 1 \\ 1 & x & x^2 \\ 1 & x^2 & x
\end{array}\rkl
  \label{eq:B4}
\end{eqnarray}
is indeed such that
\begin{equation}
  \label{eq:B5}
  \biglb|\braket{m\phpr_k}{m'_{k'}}\bigrb|^2=
\left\{\begin{array}{c@{\quad\Text{if}\quad}l}
\delta_{kk'}& m=m'\,, \\[1ex] \frac{1}{3} & m\neq m'\,,
\end{array}\right.
\end{equation}
so that each set consists of 3 orthonormal states, as it should, 
and any two different sets are complementary. 

Repeated measurements of the observables $A_m$ 
(on identically prepared \spinone\ systems) eventually determine the 
probabilities $p^{(m)}_k$ for finding their eigenstates $\ket{m_k}$.
As a consequence of their mutual complementarity, knowledge of the
probabilities for one $A_m$ contains no information whatsoever about the
probabilities for any other one.
These 12 probabilities represent 8 parameters in total, since
${p^{(m)}_0+p^{(m)}_1+p^{(m)}_2=1}$ for each of the four $A_m$s.
The statistical operator that characterizes the ensemble of identically
prepared \spinone\ systems,
\begin{equation}
  \label{eq:B6}
\rho=\sum_{m=0}^3\sum_{k=0}^2\ket{m_k}
     \lkl p^{(m)}_k-\frac{1}{4}\rkl\bra{m_k}\,,
\end{equation}
is therefore uniquely determined by the probabilities 
${p^{(m)}_k=\bra{m_k}\rho\ket{m_k}}$.
Indeed, the $A_m$s constitute a complete set of mutually complementary
observables for the \spinone\ degree of freedom.

\section{Spin-1 version of the mean king's problem}\label{sec:SaveNeck}
Accordingly, in the \spinone\ version of \emph{The King's Problem} 
either one of $A_0$, $A_1$, $A_2$, or $A_3$ is measured by the mean king's 
men, on a \spinone\ atom suitably prepared by the physicist.
Without knowing which measurement was done actually, the physicist
performs a subsequent measurement of her own, and --- after then being
told which $A_m$ was measured by the king's men --- she has to state 
correctly what they found: $0$ or $1$ or $2$.

The physicist solves the problem by first preparing a state $\ket{\Psi_0}$ in
which the given \spinone\ atom is entangled with another, auxiliary, \spinone\
atom. 
Two-atom states in which the given atom is in state
$\ket{m\phpr_k}$ and the auxiliary atom in $\ket{m'_{k'}}$ are
denoted by $\ket{m\phpr_km'_{k'}}$.
Then
\begin{eqnarray}
  \ket{\Psi_0}
&=&3^{-\half}\lkl\phbig\ket{0_00_0}+\ket{0_10_1}+\ket{0_20_2}\rkl \nonumber\\  
&=&3^{-\half}\lkl\phbig\ket{1_02_0}+\ket{1_12_1}+\ket{1_22_2}\rkl \nonumber\\  
&=&3^{-\half}\lkl\phbig\ket{2_01_0}+\ket{2_11_1}+\ket{2_21_2}\rkl \nonumber\\  
&=&3^{-\half}\lkl\phbig\ket{3_03_0}+\ket{3_13_2}+\ket{3_23_1}\rkl 
\label{eq:C1}
\end{eqnarray}
are alternative ways of writing the state she prepares.
Their equivalence is easily verified with the aid of the transformation laws
(\ref{eq:B4}).

If the king's men then measure $A_m$ on the given atom and find the value $k$,
the resulting two-atom state is the respective
$\ket{m\phpr_km'_{k'}}$ component of $\ket{\Psi_0}$.
After their measurement, there are thus all together 4 trios
of possible two-atom states.
We write them compactly as
\begin{eqnarray}
\lkl\phbig\ket{0_00_0},\ket{0_10_1},\ket{0_20_2}\rkl
&=&\lkl\phbig\ket{\Psi_0},\ket{\Psi_1},\ket{\Psi_2}\rkl{\cal U}\,,
\nonumber\\[1ex]
\lkl\phbig\ket{1_02_0},\ket{1_12_1},\ket{1_22_2}\rkl
&=&\lkl\phbig\ket{\Psi_0},\ket{\Psi_3},\ket{\Psi_4}\rkl{\cal U}\,,
\nonumber\\[1ex]
\lkl\phbig\ket{2_01_0},\ket{2_11_1},\ket{2_21_2}\rkl
&=&\lkl\phbig\ket{\Psi_0},\ket{\Psi_5},\ket{\Psi_6}\rkl{\cal U}\,,
\nonumber\\[1ex]
\lkl\phbig\ket{3_03_0},\ket{3_13_2},\ket{3_23_1}\rkl
&=&\lkl\phbig\ket{\Psi_0},\ket{\Psi_7},\ket{\Psi_8}\rkl {\cal U}\,,
\label{eq:C2}\end{eqnarray}
where the 3-rows on the right are multiplied by the unitary $3\times3$ matrix
\begin{equation}
  \label{eq:C3}
{\cal U}=\frac{1}{\sqrt{3}}\lkl\begin{array}{ccc}
1 & 1 & 1 \\ 1 & x & x^2 \\ 1 & x^2 & x \end{array}\rkl  
\end{equation}
which we met in (\ref{eq:B4}) as well.
Since the members of each trio are orthogonal to each other, the 8 two-atom
states $\ket{\Psi_1}$, \ldots, $\ket{\Psi_8}$ introduced here are orthogonal
to $\ket{\Psi_0}$ by construction. It is equally immediate that the paired
states $\ket{\Psi_{2m+1}},\ket{\Psi_{2m+2}}$ are orthogonal to each other 
for $m=0,1,2,3$. That, more generally, the orthonormality relation
\begin{equation}
  \label{eq:C4}
  \braket{\Psi_j}{\Psi_k}=\delta_{jk}\quad\text{for $j,k=0,\ldots,8$}
\end{equation}
holds also for states from different trios can be checked explicitly
(or one recognizes a special case of a more general statement 
\cite{YA+BGE:ip}).

The physicist will be able to state correctly the measurement result found by
the king's men if she can find a two-atom observable $P$ with a set of
eigenstates $\ket{P_0},\ldots,\ket{P_8}$ such that each $\ket{P_k}$ is
orthogonal to two members each of the four trios on the left of (\ref{eq:C2}). 
It is convenient to specify such states by indicating which members they are
\emph{not} orthogonal to, so that
\begin{equation}
  \label{eq:C5}
  \ket{\phbig[k_0k_1k_2k_3]}
\end{equation}
has the defining property of being orthogonal to the two-atom states that
result when measurements of $A_m$ do \emph{not} give the eigenvalue $k_m$.

In order to see how this enables her to infer the measured value, suppose the
physicist finds the two-atom system in state $\ket{\phbig[1012]}$.
She then knows that if the king's men had measured $A_0$, $A_1$, $A_2$, or
$A_3$, the respective results must have been $1$, $0$, $1$, and $2$, because
she would never find $\ket{\phbig[1012]}$ for other measurement results.

Accordingly, all that is needed to complete the solution of the \spinone\
version of the mean king's problem is the demonstration that we can have a
complete orthonormal set of two-atom states of the kind (\ref{eq:C5}).
First note that the expansion of $\ket{\phbig[k_0k_1k_2k_3]}$ in the
$\ket{\Psi_j}$ basis is given by
\begin{eqnarray}
&&\ket{\phbig[k_0k_1k_2k_3]}=\frac{1}{3}\ket{\Psi_0}\nonumber\\
&&\hspace*{3em}
+\frac{1}{3}\sum_{m=0}^3\lkl\phBig\ket{\Psi_{2m+1}}x^{k_m}
                           +\ket{\Psi_{2m+2}}x^{-k_m}\rkl\,.
  \label{eq:C6}
\end{eqnarray}
Then observe that
\begin{equation}
  \label{eq:C7}
\bigl\langle[k\phpr_0k\phpr_1k\phpr_2k\phpr_3]\big|
[k'_0k'_1k'_2k'_3]\bigr\rangle
=\frac{1}{3}\sum_{m=0}^3\delta_{k\phpr_m,k'_m}-\frac{1}{3}\,,
\end{equation}
so that two such states are orthogonal if $k\phpr_m=k'_m$ for one and only one
$m$ value.
Therefore, a possible choice of basis states for the physicist's final
measurement is
\begin{equation}
\begin{array}[b]{c}
\ket{P_0}=\ket{\phbig[0000]}\,,\quad
\ket{P_1}=\ket{\phbig[0111]}\,,\quad
\ket{P_2}=\ket{\phbig[0222]}\,,\\
\ket{P_3}=\ket{\phbig[1012]}\,,\quad
\ket{P_4}=\ket{\phbig[1120]}\,,\quad
\ket{P_5}=\ket{\phbig[1201]}\,,\\
\ket{P_6}=\ket{\phbig[2021]}\,,\quad
\ket{P_7}=\ket{\phbig[2102]}\,,\quad
\ket{P_8}=\ket{\phbig[2210]}\,.
\end{array}
  \label{eq:C8}
\end{equation}
After being told which measurement the king's men performed on the given atom,
she can then infer their measurement result correctly, and with certainty, in
the manner described above for $\ket{P_3}=\ket{\phbig[1012]}$.

\section*{Acknowledgments}
YA gratefully acknowledges the kind hospitality extended to him by Herbert
Walther during a visit at the MPI f\"ur Quantenoptik in Garching. 
BGE would like to thank Wolfgang Schleich and the University of Ulm for
financial support while part of this work was done.
YA's research is supported in part by Grant No.\ 471/98 of the Basic Research
Foundation (administered by the Israel Academy of Sciences and Humanities) and
NSF Grant No.\ PHY-9971005.


\begin{references}
\bibitem{VAA}
L. Vaidman, Y. Aharonov, and D. Z. Albert, 
\textit{How to ascertain the values of $\sigma_x$, $\sigma_y$, and $\sigma_z$
of a spin-$\frac{1}{2}$ particle},
\prl {\bf58}, 1385--1387 (1987).

\bibitem{EKW}
B.-G. Englert, Ch.\ Kurtsiefer, and H. Weinfurter, 
\textit{Universal unitary gate for single-photon 2-qubit states},
\pra (in print), LANL eprint quant-ph/0101064.

\bibitem{YA+BGE:ip}
Y. Aharonov and B.-G. Englert, 
\textit{The mean king's problem: Prime degrees of freedom},
in preparation.

\bibitem{WooFie}
W. K. Wootters and B. D. Fields,
\textit{Optimal State-Determi\-nation by Mutually Unbiased Measurements},
Ann.\ Phys.\ (NY) {\bf191}, 363--381 (1989).

\end{references}
\end{document}